\documentstyle[sprocl,epsf]{article}

\def\Journal#1#2#3#4{{#1} {\bf #2}, #3 (#4)}

\def\NPB{{\em Nucl. Phys.} B}
\def\PLB{{\em Phys. Lett.}  B}
\def\PRL{\em Phys. Rev. Lett.}
\def\PR{\em Phys. Rev.}
\def\PRD{{\em Phys. Rev.} D}
\def\AP{{\em Ann. Phys.}}

\def\ra{\rightarrow}

\def\be{\begin{equation}}
\def\ee{\end{equation}}
\def\bea{\begin{eqnarray}}
\def\eea{\end{eqnarray}}

\def\beeq{\begin{equation}}
\def\eneq{\end{equation}}
\def\beqn{\begin{eqnarray}}
\def\eeqn{\end{eqnarray}}

\def\dd{\partial}
\def\la{\raise.16ex\hbox{$\langle$}\lower.16ex\hbox{}  }
\def\ra{\, \raise.16ex\hbox{$\rangle$}\lower.16ex\hbox{} }
\def\go{\rightarrow}

\def\psibar{ \psi \kern-.65em\raise.6em\hbox{$-$} }
\def\chibar{ \chi \kern-.65em\raise.5em\hbox{$-$} }
\def\mbar{ m \kern-.75em\raise.4em\hbox{$-$}\hbox{} }
\def\Bbar{ B \kern-.73em\raise.6em\hbox{$-$}\hbox{} }

\def\ep{\epsilon}

\def\vphi{ {\varphi} }

\def\eff{{\rm eff}}

\def\LapN{{\triangle_N}}
\def\potN{{V_N}}

\def\boxit#1{$\vcenter{\hrule\hbox{\vrule\kern3pt
     \vbox{\kern3pt\hbox{#1}\kern3pt}\kern3pt\vrule}\hrule}$}
\def\bigbox#1{$\vcenter{\hrule\hbox{\vrule\kern5pt
     \vbox{\kern5pt\hbox{#1}\kern5pt}\kern5pt\vrule}\hrule}$}
\def\hugebox#1{$\vcenter{\hrule\hbox{\vrule\kern8pt
     \vbox{\kern8pt\hbox{#1}\kern8pt}\kern8pt\vrule}\hrule}$}

\begin{document}

\rightline{\small UMN-TH-1507/96}
\rightline{\small NUC-MINN-96/14-T}
\vglue .5cm

\title{Aspects of Confinement and Chiral Dynamics in 2-d QED at Finite 
Temperature\footnote{~To appear in the Proceedings of {\it DPF96} 
University of Minnesota, August 10-15}}
\author{R. RODRIGUEZ and Y. HOSOTANI}

\address{School of Physics and Astronomy, University of Minnesota\\
Minneapolis, MN 55455, USA}

\author{J.E. HETRICK}

\address{Department of Physics, University of Arizona,  Tucson,
AZ 85721, USA}

\author{S. ISO}

\address{Theory Division, KEK, Tsukuba,
Ibaraki 305, Japan}


\maketitle\abstracts{We evaluate the Polyakov loop and string tension at 
zero and finite temperature in $QED_2$. Using bozonization the problem 
is reduced to solving the Schr\"odinger equation  with a particular potential 
determined by the ground state . In the presence of two sources of 
opposite charges the vacuum angle parameter $\theta $ changes by 
$2\pi (q/e)$, independent of the number of flavors. This, in turn, 
alters the chiral condensate. Particularly, in the one 
flavor case through a simple computer algorithm, 
we explore the chiral dynamics of a heavy fermion.}

The Schwinger model, quantum electrodynamics in $1+1$ dimensions, has 
expediently been used as a metaphor of color screening 
in gauge theories.\cite{Casher}
In its massless form it is reduced to a free field theory and, consequently, 
solved exactly.\cite{Sch}
With massive fermions the theory is no longer solvable.\cite{CJS} 
Here, fractionally charged test particles are confined quite analogously 
to quark confinement in QCD.\cite{HNZ,Gross,Grignani} 

In the Matsubara formalism the model at finite temperature in 
equilibrium is equivalent to an Euclidean field theory of compact 
imaginary time $\tau .$ The strategy is to solve the model on a circle 
$S^1$ where we use the powerfull machinery of bosonization and Wick 
rotate by the replacement $L \rightarrow \beta , it \rightarrow x ,
x \rightarrow \tau .$ The Wilson line on the circle corresponds to 
the Polyakov loop in Euclidean space of compact $\tau .$

The bosonization is carried out in the interaction picture defined by free 
massless fermions: $i\gamma \partial \psi =0.$ 
\begin{figure}[t,b]
\hskip 1.8cm
\epsfxsize= 8.cm    
\epsffile[137 539 399 730]{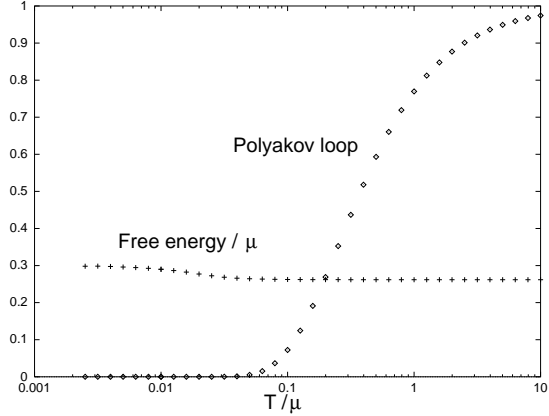}
\caption{$T$-dependence of the Polyakov loop and free energy in
the $N=3$ model with $m/\mu=.01$.  The free energy stays finite at all $T$.
\label{fig:Polyakov}}
\end{figure}
The Hamiltonian posseses a residual gauge symmetry. 
In view of this degeneracy the 
ground state of the theory is the theta vacumm, quite analogously
to QCD. 
When the fermion masses are small 
we express the vacumm wave function in terms of 
$f(\vphi_1, \cdots, \vphi_{N-1}; \theta)$.\cite{RH} 
Then, the problem is reduced to 
the solution of a Schr\"odinger equation of $N-1$ degrees of freedom. 
\beqn
&&\Big\{ -\LapN + \potN \Big\} ~ f(\vphi_1, \cdots, \vphi_{N-1})
= \ep ~ f(\vphi_1, \cdots, \vphi_{N-1}) \cr
\noalign{\kern 6pt}
&&\LapN = 
\sum_{a=1}^{N-1} {\dd^2\over \dd\vphi_a^2} 
-{2\over N-1} \sum_{a<b}^{N-1} {\dd^2\over \dd\vphi_a\dd\vphi_b} \cr
&&\potN = -
~  \sum_{a=1}^N   m_a A_a  \cos \vphi_a    \hskip .5cm   
\Big( \sum_{a=1}^N \vphi_a =\theta \Big) 
\label{Sch1}
\eeqn
where $A_a$ is determined by the boson masses $\mu_\alpha$. 
The boson masses themselves are to be determined self-consistently 
by $f(\vphi_1, \cdots, \vphi_{N-1}; \theta).$ Schematically the following
process proceeds until convergence: 
$A_a \go f(\vphi) \go \mu_\alpha \go A_a ~.$ 
\begin{figure}[t,b]
\hskip 1.2cm
\epsfxsize= 8.cm    
\epsffile[137 539 399 730]{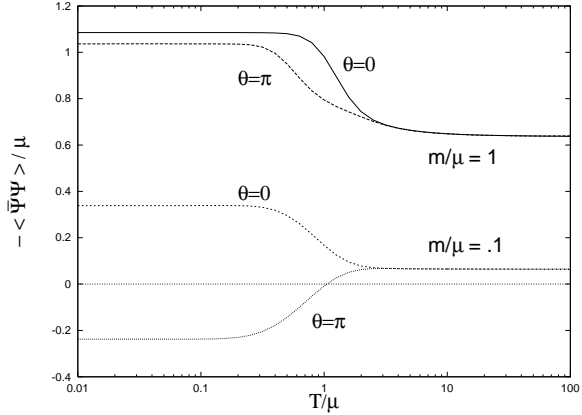}
\caption{$T$-dependence of the chiral condensates 
with fixed $m/\mu $ and $\theta $
\label{fig:cc}}
\end{figure}
We use the vacumm wave function to compute the Wilson line, from 
which we obtain the Polyakov loop.\cite{RH} Results are displayed in fig. 1.
The Polyakov loop vanishes for fractional charge due to gauge invariance.

To study confinement we pose the question: 
What is the energy of a pair of external sources of charge $q$ and
$-q$ ? This in fact, is related to the Polyakov loop correlator and
can be computed in mass perturbation theory. 
Perturbation theory cannot be employed when 
$N\ge 2$ as physical quantities are not analytic in $m$ at $T=0.$
\cite{Coleman}
Nevertheless, the change in the energy due to the introduction of external 
sources, when $m_a=m\ll \mu $ to $O(d/L)$ ($d$ is the separation), can 
be found.\cite{RH} One finds that the string tension 
$\sigma = Nm \Big\{ \la \psibar\psi\ra_{\theta_\eff}
 - \la\psibar\psi\ra_\theta \Big\} ~~,~~
\theta_\eff = \theta-2\pi q/e ~.$ 

Removing the restriction of small masses in the one flavor case, we have 
computed the chiral condensate (fig.2). Polyakov loops have also been
computed.\cite{HR}


\section*{Acknowledgments}
This work was supported in part by the U.S.\ Department of Energy
under contracts  DE-AC02-83ER-40105
(Y.H.),   DE-FG02-87ER-40328 (R.R.), and  DE-FG03-95ER-40906 (J.H.)

\end{document}